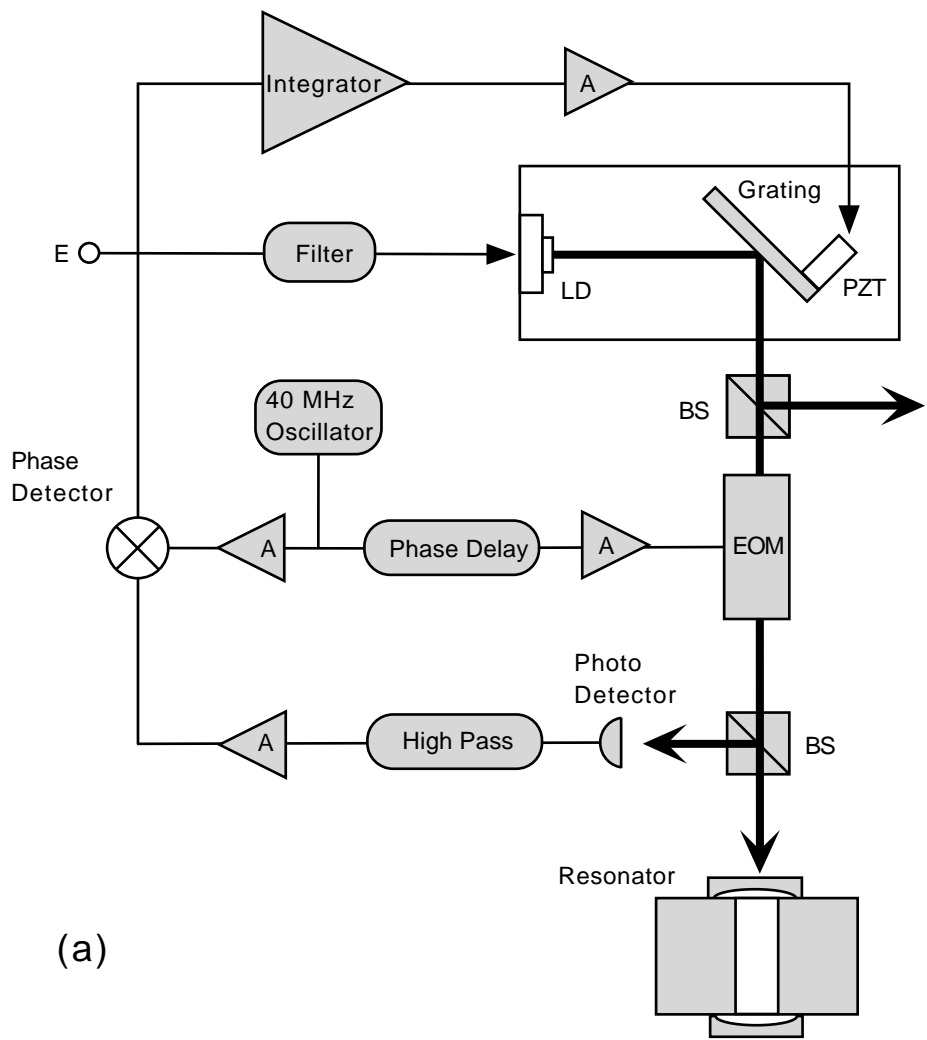

(a)

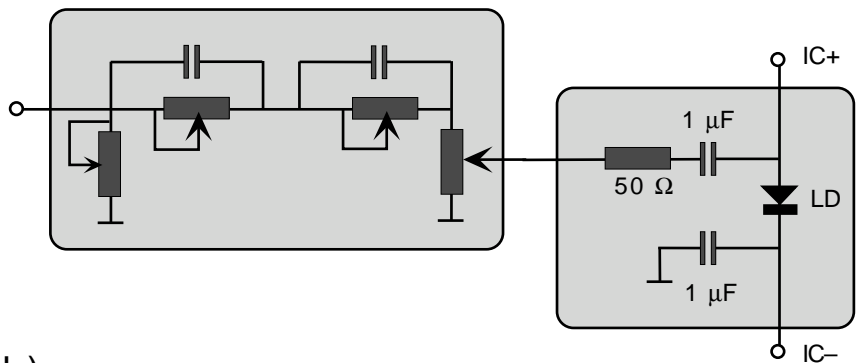

(b)

Fig.1 Schoof et al.

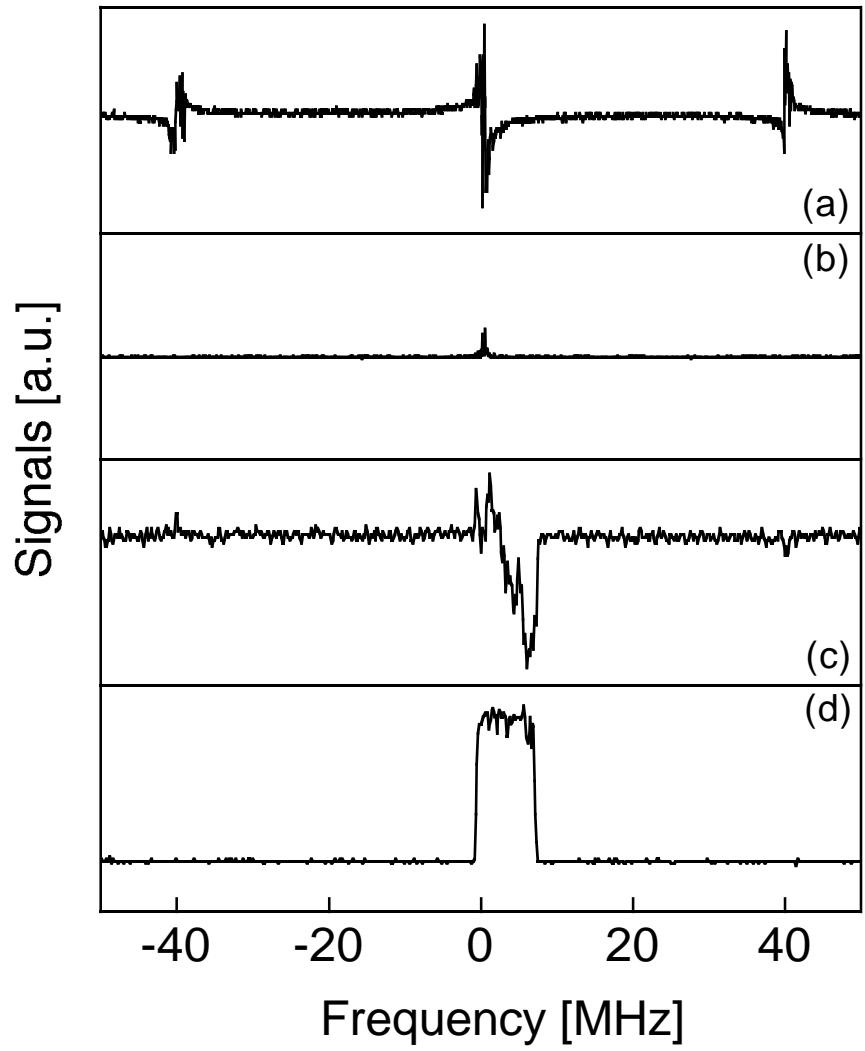

Fig.2

A. Schoof, J. Grünert, S. Ritter, A. Hemmerich

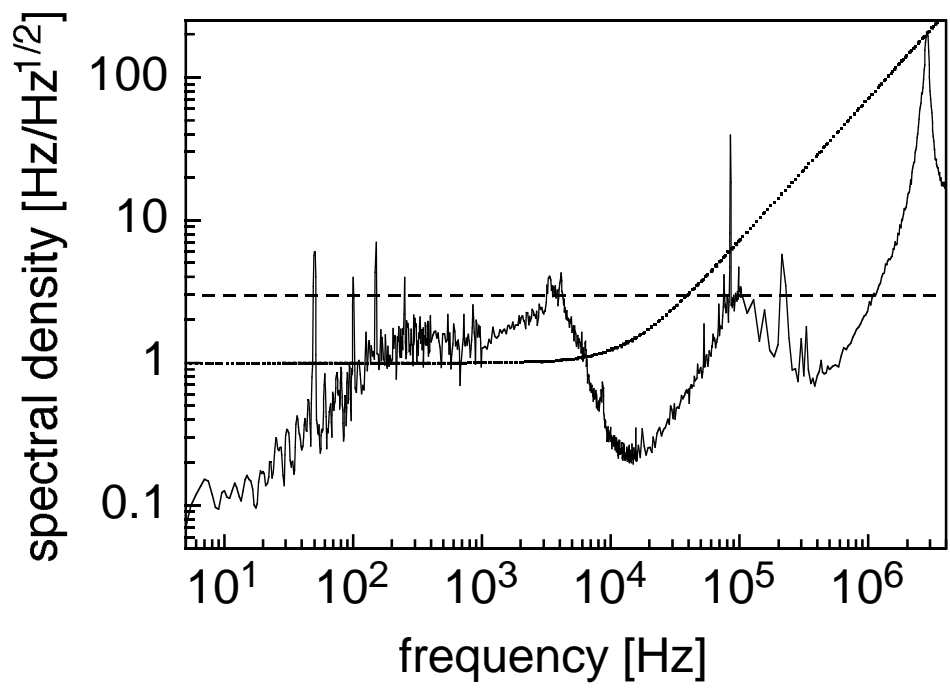

Fig.3

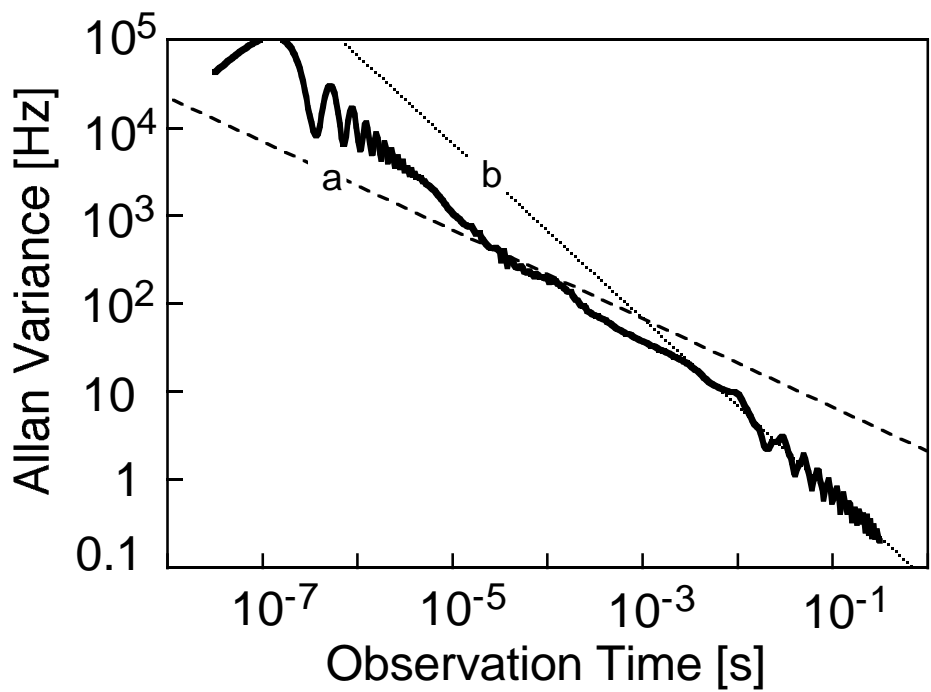

Fig.4

A. Schoof, J. Grünert, S. Ritter, A. Hemmerich

# Reducing the linewidth of a diode laser below 30 Hz by stabilization to a reference cavity with finesse above $10^5$

A. Schoof, J. Grünert, S. Ritter, and A. Hemmerich [†]

*Institut für Laserphysik, Universität Hamburg, Jungiusstraße 9,*

*D-20355  Hamburg, Germany*

An extended cavity diode laser operating in the Littrow configuration emitting near 657 nm ist stabilized via its injection current to a reference cavity with a finesse of more than $10^5$ and a corresponding resonance linewidth of 14 kHz. The laser linewidth is reduced from a few MHz to a value below 30 Hz. The compact and robust setup appears ideal for a portable optical frequency standard using the Calcium intercombination line.

During the past decade diode lasers have become one of the workhorses in modern atomic physics, quantum optics and laser spectroscopy. The desire for simple, robust and possibly portable laser sources has stimulated extensive efforts to extend their usefullness even to applications where ultimate frequency stability is a prerequisite, as e.g. for ultra high precision spectroscopy or metrology. Although linewidth reduction of diode lasers by means of reference cavities has been explored extensively [1], the kHz and sub-kHz-domains have only recently been reached. Bianchini et al. have reported a distributed Bragg reflector (DBR) laser stabilized to a reference cavity of finesse 2800, achieving better than 10 kHz mean frequency deviation (Allan variance) for observation times above 1 microsecond employing a servo bandwidth of 1 MHz [2]. A laser linewidth of about 5 kHz has been obtained in this work. Vassiliev et al. have stabilized a Littman-configured extended cavity laser to a reference cavity of finesse 5800 showing a resolution of 600 Hz in a spectroscopic application [3]. Recently, Oates et al. have used a Littman type laser in combination with an even higher finesse cavity (50 kHz line width) and could obtain a laser linewidth around 50 Hz [4].

In this paper we demonstrate a particularly robust and straight forward electronic frequency stabilization of a diode laser to a 14 kHz resonance of a Fabry-Perot resonator with a finesse of $1.1 * 10^5$. The reference cavity consists of an ultra low expansion glass (ULE) spacer and optically contacted high quality dielectric mirrors. In the present work we have placed the 10 cm long zylindrical resonator (diameter = 5 cm) on two viton rings on a slightly V-shaped aluminium platform mounted directly on the laser table with no further vibrational isolation. Employing the Pound Drever Hall (PDH) technique [5] to control the injection current of the laser diode with a servo bandwidth up to 5 MHz we obtain an emission linewidth of the locked laser below 30 Hz. To our knowledge electronically stabilized diode lasers with such narrow linewidths have not yet been reported. Our light source appears particularly suited for a portable optical frequency standard using the Calcium intercombination line.

Our experimental setup is sketched in fig.1. The light source is an extended cavity diode laser using a holographic grating in Littrow configuration (see ref. [6]), a particularly robust and compact choice at the expense of slightly less inherent frequency stability as compared to Littman type lasers. Using the Mitsubishi laser chip ML1016R-01, we typically get about 15 mW output power at 657 nm with a spectral linewidth of a few MHz integrated over a second (dominated by technical low frequency noise) and about 200 kHz within a millisecond. About 1 mW is directed to the reference resonator via an electro-optic modulator (EOM) generating frequency modulation (FM) sidebands at 40 MHz with a power of about 1 % of that of the carrier. Careful mode matching lets us couple up to 40 % of the carrier into the resonator. The light reflected from the incoupling mirror is recorded by a photo diode (EG&G FFD 040). The non-resonant sidebands and the resonant carrier produce a 40 MHz beat component which is recorded and demodulated in a phase detector (Minicircuits, RPD1). An adjustable phase delay circuit before the EOM allows to pick the desired quadrature component. The demodulated signal serves as the error signal for a proportional and an integrated feedback branch. The integrated error signal is fed to the piezo-electric transducer (PZT) adjusting the grating of the extended cavity laser. This relatively slow frequency adjusting element offers a servo bandwidth of only a few 100 Hz in combination with large frequency excursions of a few GHz. In order to extend the servo bandwidth, the error signal is high pass filtered and fed directly to the laser cathode. The adjustable high pass filter serves to compensate for frequency dependent phase lags ocurring in the diode laser chip and for the finite cavity bandwidth. The values of the variable resistors (typically with $R_{max}$ being a few hundred Ohms) and capacitors (typically a few 10 pF) in the simple passive filter circuit (sketched in fig.1 (b)) had to be adjusted according to the specific diode chip being used. The filtered error signal is AC-coupled to the laser diode as sketched in fig.1 (b).

In fig. 2 the error signal and the power transmitted through the resonator is plotted while the grating of the extended cavity laser is tuned. The upper two traces show the error signal (a) and the transmitted signal (b) when both servo loop branches are disconnected. We recognize the expected transmission resonance and the corresponding dispersive error signal. The lower two traces ((c) and (d)) show the respective signals, if the proportional servo loop is activated. When the laser frequency is tuned sufficiently close to the cavity resonance, locking to this resonance occurs and the tuning of the grating is compensated yielding a plateau-like transmitted signal in (c). The corresponding error signal (d) needed for providing this compensation is proportional to the detuning of the grating as expected. Traces (c) and (d) show that frequency excursions of more than five MHz are compensated by the fast injection current feedback loop yielding a very robust locking performance. Note also that the locked transmitted signal (d) is much larger than the unlocked one (b) because in the unlocked case the laser emission frequency during its scan is driven back and forth across resonance by noise in a time shorter than the cavity decay time of 11 microseconds.

In fig.3 we show the spectral noise density in [Hz/Hz$^{1/2}$] derived from recording the error signal for closed servo loop (at point E in fig.1) with a spectrum analyzer. In order to translate noise power correctly into frequency noise it was not sufficient to scan the unlocked laser and record the corresponding error signal because its emission spectrum was too broad for appropriately resolving the 14 kHz resonances. Thus we have extended our setup such that the laser could be locked alternatively to a second comparable cavity. An accousto-optic modulator was used for tuning across the resonance of the first cavity and the corresponding error signal was recorded with sufficient resolution. At noise frequencies above the cavity bandwith of $\nu_{cav}$=14 kHz one expects a 20 dB/decade roll-off in the conversion of noise power into frequency. This has been taken into account by calibrating the square root of the measured noise power with the factor $(1 + (\nu/\nu_{cav})^2)^{1/2}$ shown in the dotted trace in Fig.3. In the spectral density in Fig.3 three maxima can be identified. The peak at 3.8 kHz corresponds to the resonance frequency of the piezo controled grating. The relative maximum around 100 kHz results from the imperfect spectral response of the input circuitry for the injection current feedback. A refined design of the adjustable high pass filter should reduce that part of the noise. A particularly pronounced peak at 3 MHz indicates the relaxation oscillations of the fast feedback branch. The dashed line corresponds to a white frequency noise level of $S^2 = 9$ Hz$^2$/Hz which represents an upper bound of the spectral density except for the region around 3 MHz dominated by the relaxation oscillations. White frequency noise is known to yield a Lorentzian spectral power density of the laser emission with a linewidth of $\pi * S^2$ which amounts to 28 Hz in our case. [7]. The white frequency noise level of Fig.3 represents an upper bound of the noise fraction contributing to the central frequency component of the emission power spectrum of the laser while the large relaxation oscillation peak at 3 MHz adds significant frequency modulation sidebands which, however, do not disturb the spectroscopic resolution offered by the narrow band carrier. The linewidth of the carrier relevant in spectroscopic applications can be significantly lower than the upper bound value of 30 Hz quoted here. Significant improvement should be possible by optimization of the servo gain and by reduction of the technical noise below 10 kHz with the help of a vibration isolation of the reference resonator in a vacuum. The spectral noise density in fig.3 is everywhere more than 10 dB above the noise level of the electronic circuitry of about $5*10^{-3}$ Hz/Hz$^{1/2}$ which thus provides a neglible contribution to the emission linewidth. We have also measured the intensity noise of the laser with and without locking finding no significant difference. Thus we believe that the absolut stability of the laser is determined by the stability of the reference cavity.

The spectral performance of the laser can also be explored in the time domain. In fig.4 we show the Allan variance [8] $\Sigma(t)$, a time correlation function particularly suitable for analyzing noise processes, which can be calculated from the spectral noise density $S(v)$ as

$$\Sigma(t)^2 = 2 \int_0^\infty dv \, |S(v)|^2 \frac{\sin^4(\pi v t)}{(\pi v t)^2} \tag{1}$$

For long observation times above 1 ms $\Sigma(t)$ is well approximated by the expression $\Sigma_b(t) = C_b * t^{-1}$ with $C_b = 0.063$ as is indicated by the dotted line (b) in fig.4. The dashed line (a) is given by the expression $\Sigma_a(t) = C_a * t^{-1/2}$ with $C_a = 2.1$ Hz/Hz$^{1/2}$. As is directly seen from eq.1, $\Sigma_a(t)$ corresponds to white frequency noise of $S_a = \sqrt{2} \, C_a = 3$ Hz/Hz$^{1/2}$ which corresponds to a laser linewidth of $\pi * S_a^2 \approx 28$ Hz. The Allen variance exceeds this white frequency noise level only at short observation times below 15 μs as a consequence of the strong relaxation oscillation side bands.

In summary, we have stabilized a Littrow type extended cavity diode laser to a reference resonator with a finesse above $10^5$ obtaining a reduction of the emission linewidth from a few MHz down to a value below 30 Hz. In forthcoming experiments the reference cavity will be thermally and mechanically decoupled from the environment. We expect a significant reduction of the linewidth limiting technical noise below 10 kHz. Further improvements should be possible by optimized adjustment of the injection current feedback loop and by further increasing the reference cavity finesse and the servo bandwidth. Diode lasers showing a linewidth deep in the sub-Hz domain appear well within reach.

This work has been partly supported by DAAD (415 probral/bu) and DFG (DFG-He2334/2.3) and by the European human potential programme. It is a pleasure to acknowledge that this work has profited significantly from discussions with F. Riehle, J. Hall, C. Oates, and F. Cruz.

**Figure Captions**

**Fig. 1**. (a) Sketch of the experimental setup. A = amplifier, BS = beam splitter, EOM = electro-optic modulator, LD = laser diode, PZT = piezoelectric transducer. (b) Sketch of the high pass filter employed in the injection current feedback loop and the coupling to the laser diode. LD = laser diode, IC± denote the connections to the injection current driver.

**Fig. 2**. For explanation see text. Graphs (b) and (d) share the same scale for their y-axis.

**Fig. 3**. Spectral noise density in Hz/Hz$^{1/2}$ of the error signal measured at point E in the scheme of fig.1. The dashed line indicates an upper white noise level of 9 Hz$^2$/Hz. The dotted curve denotes the frequency calibration function used in the determination of the spectral noise density from the measured noise power of the error signal.

**Fig. 4**. Allan variance of spectral density of fig.3. The dotted lines indicate time dependences $t^{-1/2}$ (a) and $t^{-1}$ (b).